\documentclass{appolb}
\usepackage{epsfig}

\newcommand{\msun}{M_{\odot}}

\begin{document}
\title{Phase Transitions in Dense Baryonic Matter and Cooling of
  Rotating Neutron Stars%
  \thanks{This work was supported by the National Science Foundation
    under Grant PHY--0854699.}%
} \author{Fridolin Weber and Rodrigo Negreiros
  \address{Department of Physics, San Diego State University, 5500
    Campanile Drive \\ San Diego, CA 92182, USA} 
}
\maketitle
\begin{abstract}
  New astrophysical instruments such as skA (square kilometer Array)
  and IXO (formerly Constellation X) promise the discovery of tens of
  thousands of new isolated rotating neutron stars (pulsars), neutron
  stars in low-mass X-ray binaries (LMXBs), anomalous X-ray pulsars
  (AXPs), and soft gamma repeaters (SGRs).  Many of these neutron
  stars will experience dramatic density changes over their active
  lifetimes, driven by either stellar spin-up or spin-down, which may
  trigger phase transitions in their dense baryonic cores. More than
  that, accretion of matter onto neutron stars in LMXBs is believed to
  cause pycno-nuclear fusion reactions in the inner crusts of neutron
  stars. The associated reaction rates may be drastically altered if
  strange quark matter would be absolutely stable. This paper outlines
  the investigative steps that need to be performed in order to
  explore the thermal response of neutron stars to rotationally-driven
  phase transitions in their cores as well as to nuclear burning
  scenarios in their crusts. Such research complements the exploration
  of the phase diagram of dense baryonic matter through particle
  collider experiments, as performed at RHIC in the USA and as planned
  at the future Facility for Antiproton and Ion Research (FAIR) in
  Darmstadt, Germany.
\end{abstract}
  
\section{Introduction}

\begin{figure}[tb] 
\begin{center}
\includegraphics*[width=0.60\textwidth,angle=0]{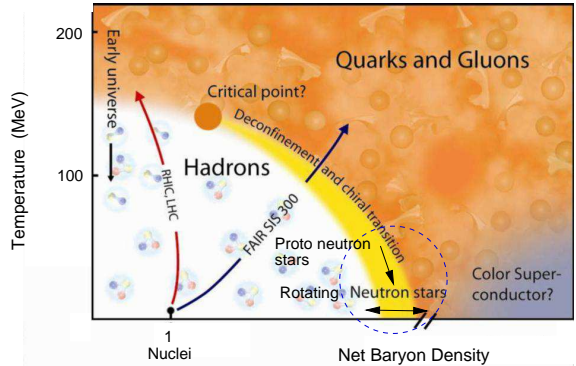}
\caption[]{Schematic phase diagram of strongly interacting matter
  \cite{FAIR09:phasediagram}. The dashed circle indicates the portion
  of the phase diagram that can be explored by studying either hot and
  newly formed proto neutron stars (arrow pointing downwards), or cold
  rotating neutron stars (vertical double-headed arrow) such as
  millisecond pulsars and neutron stars in low-mass X-ray binaries
  (LMXBs).}
\label{fig:FAIR}
\end{center}
\end{figure}

On the Earth, particle collider experiments enable physicists to cast
a brief glance at the properties of ultra-dense and hot baryonic
matter (Fig. \ref{fig:FAIR}). On the other hand, it is estimated that
galaxies like our Milky Way contain between $10^8$ and $10^{10}$
collapsed stars known as neutron stars, which harbor ultra-dense
matter permanently in their cores. This key feature together with the
unprecedented progress in observational astrophysics, which is
expected to be excelled by future observatories such as the
square kilometer Array (skA) and Constellation-X, make neutron stars
superb astrophysical laboratories for a wide range of physical
studies. 
These studies concern nuclear fusion processes on the stellar surface,
pycnonuclear reactions in electron degenerate matter at sub-nuclear
densities, and the possible formation of boson condensates and other
novel states of baryonic matter--like color superconducting quark
matter--at super-nuclear densities.  (For overviews, see, for instance
\cite{glen97:book,weber99:book,blaschke01:trento,lattimer01:a,rajagopal01:a,%
  weber05:a,klahn06:a_short,page06:a,sedrakian07:a,alford08:a}.) More
than that, there is the very intriguing theoretical suggestion that
strange quark matter could be more stable than atomic nuclei, known as
the strange quark matter hypothesis, in which case neutron stars
should be largely composed of pure strange quark matter
\cite{alcock86:a,alcock88:a,madsen98:b}. If quark matter exists in
neutron stars it ought to be a color superconductor
\cite{rajagopal01:a,alford01:a,alford98:a,rapp98+99:a}. Other testable
implications of the strange quark matter hypothesis concern the
possible existence of a new class of white dwarfs, known as strange
white dwarfs \cite{glen92:crust,glen94:a}, and the drastic alteration
of heavy-ion reaction rates in the deep crustal layers of neutron
stars \cite{golf09:a}, if strange quark matter nuggets should be
present in these layers.

\section{Phase transitions in the cores of neutron stars}
\label{sec:phasetr}

Rotating neutron stars are called pulsars. Three distinct classes of
pulsars are currently known. These are (1) rotation powered pulsars,
where the loss of rotational energy powers the emitted
electromagnetic radiation, (2) accretion-powered (X-ray) pulsars,
where the gravitational potential energy of the matter accreted from a
low-mass companion is the energy source, and (3) magnetars, where the
decay of a ultra-strong magnetic field powers the radiation.
Depending on star mass and rotational frequency, the matter in the
core regions of neutron stars may be compressed to densities that are
up to an order of magnitude greater than the density of ordinary
atomic nuclei. This extreme compression provides a high-pressure
environment in which numerous subatomic particle processes are likely
to take place
\cite{glen97:book,weber99:book,blaschke01:trento,sedrakian07:a}.  The
most spectacular ones stretch from the generation of hyperons and
baryon resonances ($\Sigma, \Lambda, \Xi, \Delta$), to quark ($u, d,
s$) deconfinement, to the formation of boson condensates ($\pi^-, K^-,
H$)
\cite{glen97:book,weber99:book,lattimer01:a,weber05:a,page06:a,sedrakian07:a}.
Rapid rotation can change the structure and composition of neutron
stars substantially, depending on the equation of state of
ultra-dense baryonic matter.
\begin{figure}[tb]
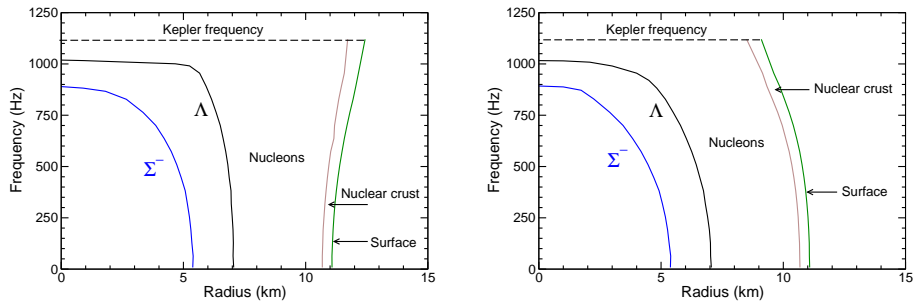

\begin{center}
\begin{tabular}{cc}
\includegraphics[width=0.45\textwidth]{1.40_eq_bonn.eps} ~~&
\includegraphics[width=0.45\textwidth]{1.40_po_bonn.eps}
\end{tabular} 
\caption{Composition of a rotating neutron star in equatorial
  direction (left panel) and polar direction (right panel)
  \cite{weber07:HYP}. The star's mass at zero rotation is $1.40\,
  \msun$.}
\label{fig:bonn_eq_po_1.40}
\end{center}
\end{figure}
The most rapidly rotating, currently known neutron star is pulsar PSR
J1748-2446ad, which rotates at a period of 1.39~ms (which corresponds
to a rotational frequency of 719~Hz) \cite{hessels06:a}. It is
followed by PSRs B1937+21 \cite{backer82:a} and B1957+20
\cite{fruchter88:a} whose rotational periods are 1.58 ms (633~Hz) and
1.61~ms (621~Hz), respectively. Finally, we mention the recent
discovery of X-ray burst oscillations from the neutron star X-ray
transient XTE J1739--285 \cite{kaaret06:a,bejger07:a}, which could
suggest that XTE J1739--285 contains an ultrafast neutron star
rotating 1122 Hz.
\begin{figure}[tb]
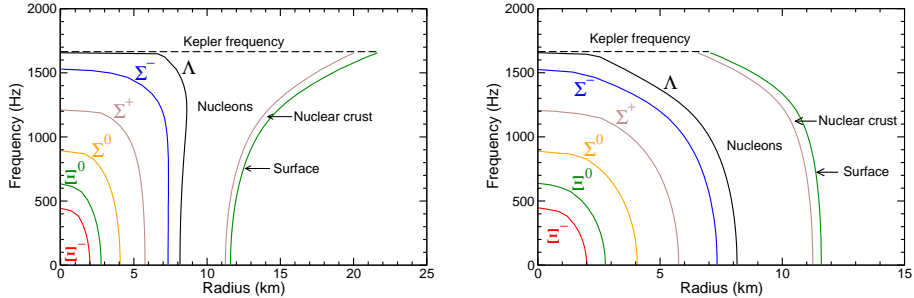

\begin{center}
\begin{tabular}{cc}
\includegraphics[width=0.45\textwidth]{1.70_eq_bonn.eps} ~~&
\includegraphics[width=0.45\textwidth]{1.70_po_bonn.eps}
\end{tabular} 
\caption{Same as Figure \ref{fig:bonn_eq_po_1.40}, but for
  a non-rotating stellar mass of $1.70\, \msun$ \cite{weber07:a}.}
\label{fig:bonn_eq_po_1.70}
\end{center}
\end{figure}
Rotating neutron stars appear as much better probes for the structure
of dense baryonic matter than non-rotating neutron stars, primarily
because the particle compositions in rotating neutron stars are not
frozen in, as it is the case for non-rotating neutron stars, but are
varying with time. The associated density changes can be as large 60\%
\cite{weber99:book} in neutron stars in binary stellar systems (e.g.,
LMXBs), which are being spun up to high rotational frequencies, or
isolated rotating neutron stars (e.g. isolated millisecond pulsars)
which are spinning down to low frequencies because of the emission of
gravitational radiation, electromagnetic dipole radiation, and a wind
of electron-positron pairs. As an example, we show the
rotationally-driven restructuring effects in the cores of standard
neutron stars in Figs.\ \ref{fig:bonn_eq_po_1.40} and
\ref{fig:bonn_eq_po_1.70}. Qualitatively similar restructuring effects
were obtained for neutron stars containing quark matter
\cite{weber99:book,weber05:a}.

\section{Thermal evolution of rotating neutron stars}
\label{ssec:nutshell}

\begin{figure}[tb] 
\begin{center}
\includegraphics*[width=0.60\textwidth,angle=0]{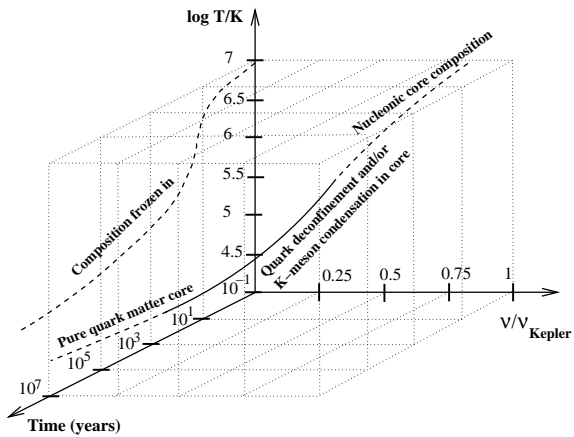}
\caption[]{Schematic difference of the cooling behavior of rotating and
  non-rotating neutron stars. The composition is frozen in in non-rotating
  neutron stars, but varies strongly with rotational frequency, $\nu$, in
  rotating neutron stars during spin-down/up. This restructuring effect can
  alter the cooling behavior of rotating neutron stars significantly. Further
  differences arise from the different geometry (metric) of rotating stars and
  from anisotropic heat transport.}
\label{fig:schematic-cooling}
\end{center}
\end{figure}

\begin{figure}[tb] 
\begin{center}
\includegraphics*[width=0.65\textwidth,angle=0]{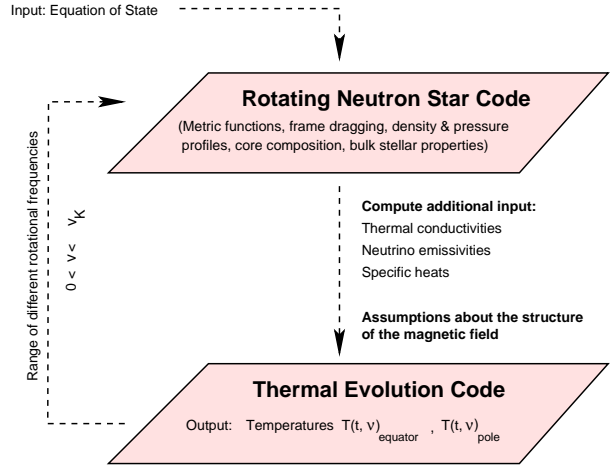}
\caption[]{Combination of numerical codes required to perform thermal
  evolution studies of rotating neutron stars.}
\label{fig:schematic-codes}
\end{center}
\end{figure}

Figure \ref{fig:schematic-cooling} illustrates schematically the
differences of the cooling behavior of rotating and non-rotating
neutron stars.
Computing the thermal evolution of rotating neutron stars, however, is
considerably more complicated than computing the cooling behavior of
non-rotating neutron stars, for several reasons
\cite{weber99:book}. First, stellar rotation requires solving
Einstein's field equations for rotationally deformed fluid
distributions, which renders the problem 2-dimensional. Second, the
general relativistic frame dragging (Lense-Thirring) leads to the
appearance of additional non-linear field equation. Third, the
determination of the general relativistic Kepler frequency, which sets
an absolute limit on stable rapid rotation, leads to an additional
self-consistency condition \cite{weber99:book}. Fourth, the thermal
transport equations need to be solved for general relativistic,
non-spherical fluids that may experience anisotropic heat
transport. Because of all these complications, a fully self-consistent
general relativistic treatment of the cooling of rotating neutron
stars has not been achieved yet. However, first steps toward this goal
were made in \cite{stejner08:a}. 
The basic cooling features of a neutron star are easily grasped by
considering the energy conservation relation of the star in the
Newtonian limit \cite{page05:a}. This equation is given by
\begin{eqnarray}
dE_{\rm th}/dt= C_V dT/dt = -L_\nu - L_\gamma + H \, , 
\label{eq:cooling1}
\end{eqnarray}
where $E_{\rm th}$ is the thermal energy content of a neutron star,
$T$ its internal temperature, and $C_V$ its total specific heat. The
energy sinks are the total neutrino luminosity, $L_\nu$, and the
surface photon luminosity, $L_\gamma$. The source term $H$ includes
all possible heating mechanisms \cite{page05:a}, which, for instance,
convert magnetic or rotational energy into heat. The dominant
contributions to $C_V$ come from the core whose constituents are
leptons, baryons, boson condensates and possibly deconfined
superconducting quarks.  When baryons and quarks become paired, their
contribution to $C_V$ is strongly suppressed at temperatures smaller
than the critical temperatures associated with these pairing
phases. The crustal contribution is in principle dominated by the free
neutrons in the inner stellar crust but, since these are extensively
paired, practically only the nuclear lattice and electrons
contribute. Extensive baryon, and quark, pairing can thus
significantly reduce $C_V$. In order to derive the general
relativistic version of Eq.\ (\ref{eq:cooling1}) for rotating stars,
one needs to solve Einstein's field equations using the metric of a
rotationally deformed fluid \cite{weber99:book},
\begin{eqnarray}
ds^2 = - e^{2 \nu} dt^2 + e^{2 \phi} (d\varphi - N^\varphi dt)^2
+ e^{2 \omega} (dr^2 + r^2 d\theta^2) \, ,
\label{eq:3.1}
\end{eqnarray}
where $e^{2\phi} \equiv e^{2(\alpha + \beta)} r^2 \sin^2\theta$ and $e^{2
  \omega} \equiv e^{2(\alpha-\beta)}$.  The quantities $\nu$, $\phi$ and
$\omega$ denote metric functions, and $N^\varphi$ accounts for frame dragging
caused by the rotating fluid. All these functions are to be computed
self-consistently from Einstein's field equation, $G^{\bar\alpha \bar\beta} =
8 \pi T^{\bar\alpha \bar\beta}$, where $T^{\bar\alpha \bar\beta}$ denotes the
fluid's energy momentum tensor. In the case of neutron star cooling,
the equations which describe the conservation of energy and momentum in a
co-moving reference frame follow from
$T^{\bar\alpha \bar\beta}{}_{;\bar\beta} = q^{\bar\alpha}  \, ,$
where
\begin{eqnarray}
T^{\bar\alpha\bar\beta} =
\left( \begin{array}{cccc}
J                  &H_r         &H_\theta        &H_\varphi  \\
H_r                &J/3          &0             &0   \\   
H_\theta            &0           &J/3           &0   \\   
H_\varphi           &0           &0              &J/3  \\
\end{array} \right)  \, , \qquad
q^{\bar\alpha} = 
\left( \begin{array}{c}
- \epsilon                 \\
-H_r / \Lambda              \\   
-H_\theta / \Lambda              \\   
-H_\varphi / \Lambda              \\   
\end{array} \right) \, .
\label{eq:3.10+3.11}
\end{eqnarray}
The quantity $q^{\bar\alpha}$ describes the interaction of photons with
matter, $J=a T^4$ is the energy density of the photon field,
$H_i$ is the i$^{\rm th}$ component of the heat flux, $\epsilon$ the emitted
energy, and $\Lambda$ the mean free path of photons.  From Eq.\
(\ref{eq:3.10+3.11}) one obtains
\begin{eqnarray}
q^{\bar\alpha} = e^\beta_{\bar\beta} \, T^{\bar\alpha \bar\beta}{}_{;\beta} 
+ \gamma^{\bar\alpha}_{\bar\gamma\bar\beta} \, T^{\bar\gamma \bar\beta} 
+ \gamma^{\bar\beta}_{\bar\gamma\bar\beta} \, T^{\bar\alpha \bar\gamma}{} \, ,
\label{eq:3.13}
\end{eqnarray}
where the $\gamma$'s denote the Ricci rotation coefficients given by
$\gamma^{\bar\alpha}_{\bar\beta\bar\gamma} = e^{\beta}_{\bar\beta} \,
e^{\bar\alpha}_{\alpha} e^{\alpha}_{\bar\gamma;\beta}$. Combining Eqs.\
(\ref{eq:3.10+3.11}) and (\ref{eq:3.13}) leads to the general
relativistic transport equations
\begin{eqnarray}
\partial_r \tilde H_{\bar r} + {1 \over r} \partial_\theta
\tilde H_{\bar\theta} &=& - r \, e^{\phi + 2\omega}
\left( {1 \over \Gamma} e^{2\nu} \epsilon + \Gamma C_V \partial_t 
\tilde T\right) \nonumber \\ & & - r \, \Gamma U e^{\nu+2\phi+\omega} 
\left( \partial_r \Omega + {1 \over r} \partial_\theta \Omega \right) ,
\label{eq:3.23} \\
\partial_r \tilde T &=& - {1 \over{r \kappa}} e^{\nu -\phi} \tilde H_{\bar r}
- \Gamma^2 U e^{-\nu + \phi} \, \tilde T \partial_r \Omega \, ,
\label{eq:3.24} \\
{1 \over r} \partial_\theta \tilde T &=& - {1 \over{r \kappa}}
e^{-\nu -\phi} \tilde H_{\bar \theta}
- \Gamma^2 U e^{-\nu + \phi} \, \tilde T {1 \over r} \partial_\theta \Omega \, , 
\label{eq:3.25} \\
\Gamma U \partial_t \tilde T &=& - {1 \over{r \kappa}}
e^{-\omega -\phi} \tilde H_{\bar \varphi} \, ,
\label{eq:3.26}
\end{eqnarray} 
where $\tilde H_i \equiv r e^{2\nu+\phi+\omega} H_i / \Gamma$, $\tilde T
\equiv e^\nu T / \Gamma$, and the Lorentz factor $\Gamma \equiv (1 -
U^2)^{-1/2}$. In the case of uniform rotation, $\Omega=$const, Eqs.\
(\ref{eq:3.23}) to (\ref{eq:3.26})
reduce to
\begin{eqnarray}
\partial_r \tilde H_{\bar r} + {1 \over r} \partial_\theta
\tilde H_{\bar\theta} &=& - r \, e^{\phi + 2(\alpha-\beta)}
\left( {1 \over \Gamma} e^{2\nu} \epsilon + \Gamma C_V \partial_t 
\tilde T\right)  ,
\label{eq:3.29} \\
\partial_r \tilde T &=& - {1 \over{r \kappa}} e^{-\nu -\phi} \tilde H_{\bar r}
\, ,
\label{eq:3.30} \\
{1 \over r} \partial_\theta \tilde T &=& - {1 \over{r \kappa}}
e^{-\nu -\phi} \tilde H_{\bar \theta} \, .
\label{eq:3.31} 
\end{eqnarray} 
The standard cooling equations of spherically symmetric, non-rotating neutron
stars are obtained from Eqs.\ (\ref{eq:3.29}) to (\ref{eq:3.31}) for $\Omega=
0$ and $\partial_\theta \tilde T=0$ \cite{weber99:book}. This project aims at
solving Eqs.\ (\ref{eq:3.23}) to (\ref{eq:3.25}) for the temperature
distribution $T(r,\theta;t)$ of non-spherical, (possibly differentially)
rotating neutron stars.  The boundary condition are given by defining $\tilde
H_{\bar r}$ at $r=0, R$, and $\tilde H_{\bar \theta}$ at $\theta=0, \pi/2$ and
at $r=R$, with $R$ denoting the stellar radius. The choice for the star's
initial temperature, $T(r,\theta;t=0)$, is typically chosen as $\tilde T
\equiv 10^{11}$~K.  Equations (\ref{eq:3.24}) and (\ref{eq:3.25}) can be
solved for $\tilde H_{\bar r}$ and $\tilde H_{\bar \theta}$, differentiated
with respect to $\partial_r$ and $\partial_\theta$, respectively, and
substituted into Eq.\ (\ref{eq:3.23}), which leads to the following parabolic
differential equation,
\begin{eqnarray}
  \partial_t \tilde T &=& - \, {1 \over{\Gamma^2}} e^{2 \nu} {\epsilon \over{
      C_V}} - r \sin \theta \, U e^{\nu+\gamma-\xi} {1 \over{C_V}}
\left( \partial_r \Omega + {1 \over r} \partial_\theta \Omega \right) 
\nonumber \\
&& + {1 \over{r^2 \sin\theta}} {1 \over \Gamma} e^{3\nu-\gamma-2\xi}
{1 \over {C_V}} \Bigl( \partial_r \left( r^2 \kappa \sin \theta \, e^\gamma
\left( \partial_r \tilde T + \Gamma^2 U e^{-2\nu+\gamma} \tilde T \partial_r
  \Omega 
\right) \right)  \nonumber \\
&& + {1 \over{r^2}} \partial_\theta \left( r^2 \kappa \sin \theta \, e^\gamma
\left( \partial_\theta \tilde T + \Gamma^2 U e^{-2\nu+\gamma} \tilde
  T \partial_\theta \Omega \right) \right) \Bigr) \, ,
\label{eq:3.49}
\end{eqnarray}
with the definitions $r \sin\theta e^{-\nu+\gamma} = e^\phi$ and
$e^{-\nu+\xi} = e^{\alpha-\beta}$. It is this differential equation
that needs be solved numerically in combination with a general
relativistic stellar rotation code, as illustrated schematically in
Fig.\ \ref{fig:schematic-codes}.  A 2-dimensional, general
relativistic stellar rotation code is to be used to determine the
metric functions, frame dragging frequency, pressure and density
gradients, and particle compositions of a deformed neutron star as a
function of rotational frequency. Depending on the cooling channels
that are active at a given rotational frequency, the numerical outcome
obtained from the rotation code then serves as an input for the
general relativistic 2-dimensional thermal evolution code, which is
used to determine the luminosity, and thus the surface temperature, of
a deformed rotating neutron star. As outlined in Sect.\
\ref{sec:phasetr}, because of stellar spin-down or spin-up, the
density in such stars may change dramatically so that new cooling
channels open up (or close) with time (Fig.\
\ref{fig:schematic-cooling}). The new cooling channels imply neutrino
emissivities, heat capacities, and thermal conductivities that are
different from the original ones, thus altering the stars thermal
response. It is this response in the thermal behavior of rotating
neutron stars that carries information about the properties of the
matter in the dense baryonic stellar cores and the deep crustal
layers.  Several key research activities that can be addressed with
these codes are briefly summarized next:

\begin{itemize}

\item Study the impact of latent heat, released by rotation-driven
  quark-hadron phase transitions in the cores of isolated rotating
  neutron stars, on the thermal evolution of such objects
  \cite{stejner08:a}.

\item Simulate the impact of latent heat release on the thermal
  evolution of accreting neutron stars in LMXBs.  Because of the
  spin-up of such neutron stars, these objects becomes progressively
  more decompressed with time, which may lead to drastic changes in
  their core compositions, causing the destruction of certain states
  ($K-$ condensate, quark matter) of dense baryonic matter that were
  initially present at slow stellar rotation rates. This modifies the
  star's thermal evolution and may lead to signals indicative of the
  existence of phase transitions in the star.

\item Study the impact of energy released by pycnonuclear burning of
  matter in the inner crusts of neutron stars. In particular, the
  dependence of the stellar surface temperature on the possible
  presence of strange
\begin{figure}[htb]
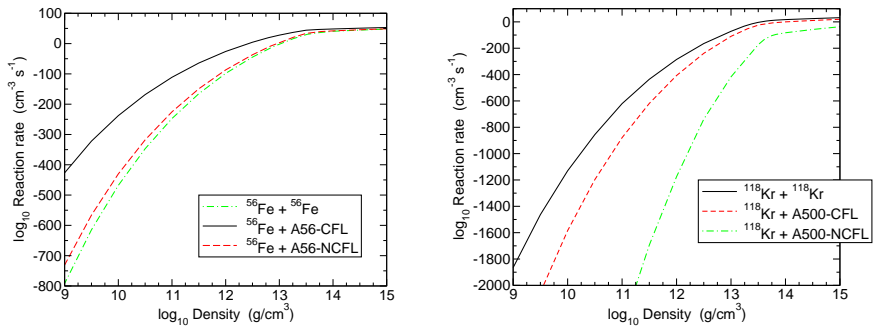

\begin{center}
\begin{tabular}{cc}
\includegraphics[width=0.41\textwidth]{Fe56SQM.eps} ~~&
\includegraphics[width=0.45\textwidth]{Kr118LargeSQM.eps}
\end{tabular}
\caption{Nuclear reaction rates for $^{56}$Fe, $^{118}$Kr, and strange
  quark matter nuggets with baryon numbers of 56 and 500
  \cite{golf09:a}.}
\label{fig:pycno}
\end{center}
\end{figure}
quark matter nuggets in the inner crusts of neutron stars can be
explored. As shown in Fig.\ \ref{fig:pycno}, the nuclear reactions
rates are strongly modified if strange quark matter nuggets would be
present in the crusts of neutron stars \cite{golf09:a}.

\item The thermal evolution of neutron stars undergoing episodes of
  intense accretion, alternated by long periods of quiescence, i.e.\
  soft X-ray transients (SXRTs) such as SAX J1808.4--3658, can be
  explored.  This neutron star is the first known accretion powered
  millisecond pulsars rotating rapidly (at 2.5~ms, 400~Hz). At such
  small spin periods certain phases of dense matter may have been spun
  out of the core, altering the neutrino loss.  The quiescent emission
  of SAX J1808.4--3658, therefore, opens the possibility of using this
  object as a tool for probing the core compositions of neutron stars
  \cite{page05:a,colpi01:a}.  Another, and still poorly explored,
  aspect of SXRTs is their short time-scale thermal response to the
  accretion phases.  This is an aspect of the problem from which
  extremely important information about both the structure of the
  neutron star crust and the thermal state of its core can be
  obtained \cite{ushormirsky01:a,ebrown02:a}, and about which
  intriguing observational results were found \cite{wijnand04:a}.

\item  Another interesting problem concerns the thermal evolution of neutron
stars immediately after they went through the proto-neutron star phase. Of
particular interest is the crystallization behavior of such matter and the
occurrence of thermoelectric instabilities at the core--crust boundary. 

\end{itemize} 



\end{document}